# ENHANCE THE DETECTION OF DOS AND BRUTE FORCE ATTACKS WITHIN THE MQTT ENVIRONMENT THROUGH FEATURE ENGINEERING AND EMPLOYING AN ENSEMBLE TECHNIQUE

Abdulelah Al Hanif and Mohammad Ilyas

Department of Electrical Engineering and Computer Science, Florida Atlantic University, Boca Raton, FL, USA

*ABSTRACT*

*The rapid development of the Internet of Things (IoT) environment has introduced unprecedented levels of connectivity and automation. The Message Queuing Telemetry Transport (MQTT) protocol has become recognized in IoT applications due to its lightweight and efficient features; however, this simplicity also renders MQTT vulnerable to multiple attacks that can be launched against the protocol, including denial of service (DoS) and brute-force attacks. This study aims to improve the detection of intrusion DoS and brute-force attacks in an MQTT traffic intrusion detection system (IDS). Our approach utilizes the MQTT dataset for model training by employing effective feature engineering and ensemble learning techniques. Following our analysis and comparison, we identified the top 10 features demonstrating the highest effectiveness, leading to improved model accuracy. We used supervised machine learning models, including Random Forest, Decision Trees, k-Nearest Neighbors, and XGBoost, in combination with ensemble classifiers. Stacking, voting, and bagging ensembles utilize these four supervised machine-learning methods to combine models. This study's results illustrate the proposed technique's efficacy in enhancing the accuracy of detecting DoS and brute-force attacks in MQTT traffic. Stacking and voting classifiers achieved the highest accuracy of 0.9538. Our approach outperforms the most recent study that utilized the same dataset.*

## 1. INTRODUCTION

The Internet of Things (IoT) has dramatically revolutionized our lives by providing and facilitating uninterrupted connectivity. The IoT is widely used because of its capability to interconnect and exchange data among devices and systems across various domains, including healthcare and smart homes. The number of IoT devices is expected to reach 38.6 billion by 2025 and is predicted to increase to 50 billion by 2030 [1]. However, the extensive utilization of IoT devices introduces a wide variety of cybersecurity challenges. Vulnerabilities in IoT networks can be exploited by somebody not permitted access to sensitive information, causing damage to the network. As a consequence, it is essential to implement robust security for the purpose of keeping a constant monitoring system and rapidly identifying and preventing any possible malicious activities on the network traffic of IoT environments.

In the IoT network environment, the communication between other devices comes through specific protocols to provide various services. These protocols establish rules for communication among devices and ensure efficient and accurate data exchange, such as MQTT, CoAP (Constrained Application Protocol), and HTTP (Hypertext Transfer Protocol) [2]. In particular, the MQTT protocol has been widely implemented in various domains of the IoT ecosystem, such as smart homes, industrial applications, healthcare, and agriculture [3]. The MQTT protocol provides lightweight, minimized packet loss, efficient communication, low memory usage, and the ability to facilitate communication with limited bandwidth [4]. The MQTT protocol is specifically designed to be lightweight for machine-to-machine communication. The MQTT architecture consists of a publisher, a subscriber, and a broker. The publisher is a transmitting device that releases the specific topic message through the broker, the broker is the central node as a server that works to facilitate communication, and the subscriber is a device that receives the specific topic message from the broker [5]. As the MQTT protocol is considered one of the foundational frameworks in IoT applications, it needs to implement robust security measures to protect against cybersecurity vulnerabilities.

In general, IoT networks are naturally vulnerable to potential risks for several types of data-related threats, unauthorized access, compromised availability, and compatibility threats, such as denial-of-service (DoS), brute force attacks, etc. Attackers typically aim to discover and exploit weaknesses and limitations within the MQTT protocol and others. In recent years, machine learning (ML) has become a valuable technique in IoT security [6]. DoS and brute force attacks are particularly hazardous among these threats. DoS attacks aim to disrupt the regular operations of IoT devices by requesting massive amounts of flows illegitimately, which makes them unavailable to authorized users and causes them to deplete essential resources such as bandwidth and hardware [7]. Brute force attacks are common in IoT security networks and can cause significant harm. Brute force attacks attempt to get unauthorized access to different combinations of usernames and passwords to obtain illegal access to devices or systems [4].

The intrusion detection system (IDS) is considered one of the methods that can be used to improve a network's security. The IDS is a valuable software application tool or monitoring device designed to detect attacks by continuously monitoring network activity and identifying possible security threats [8]. It has the ability to identify both malicious assaults and regular data and facilitate the detection of security in real time, improving the overall security of the system and network.

To enhance intrusion detection, developers and researchers need to provide advanced methods to identify impacted IoT devices. In recent years, ML has become a valuable technique in the domain of IoT security. The rapid advancements in ML and artificial intelligence (AI) algorithms currently enable network monitoring and the identification of upcoming cyber-attacks. ML and deep learning (DL) methods have become the leading methods in this domain because of their capabilities to exceed traditional IDSs [9]. ML algorithms are used to train models capable of distinguishing between regular and anomalous activities inside the system IoT environment [10]. ML methods offer an adequate foundation for developing and enhancing the security of networks and IoT devices [11].

The MQTT protocol is susceptible to attacks due to its original focus on simplicity and efficiency. It is crucial to develop IDS with high accuracy to effectively identify attacks against the MQTT protocol. While machine learning (ML) encounters specific difficulties in achieving high accuracy and performance, ensemble methods offer different techniques to address these problems [12]. The idea of ensemble methods is to combine multiple ML models by training on various types of data to enhance the overall accuracy and performance of the models, which results in better decisions than a base model [13]. Many researchers have found ensemble methods have effectively improved the performance of intrusion detection systems for IoT environments and MQTT protocol [14]. Ensemble machine learning approaches and feature engineering techniques are employed to build an intelligent IDS for MQTT traffic. This IDS effectively detects DoS and brute force attacks, issuing warnings or implementing measures to halt the attack. The incorporation of feature selection within the feature engineering process is crucial, as it substantially enhances the overall effectiveness of ensemble machine learning algorithms. Several ensemble methods are available for IoT security, the most commonly used aggregators being bagging, boosting, voting, and stacking [15]. These methods effectively protect against IoT cyber-attacks.

This research introduces an IDS for MQTT traffic that uses feature engineering and an ensemble learning algorithm to enhance and identify malicious activities of DoS and brute force attacks. The main contributions of this article are as follows:

- An approach to feature subset selection techniques involving K-Best feature, Pearson Coefficient Correlation (PCC), and Principal Component Analysis (PCA) is utilized to identify essential features in the dataset. After comparing and analyzing all the features, we opted for the top 10 due to their significant impact on improving model performance and accuracy.
- We evaluate multiple ML methods to choose the most suitable models for building an effective IDS to detect DoS and brute-force attacks on IoT MQTT traffic. This includes evaluating four supervised models—Decision Tree (DT), Random Forest (RF), k-nearest Neighbor (KNN), and Xtreme Gradient Boosting (XGBoost)—employing data balancing and cross-validation. Subsequently, we utilize three ensemble methods—stacking, voting, and bagging—to combine the four supervised ML models and enhance attack detection performance. Our evaluation performance includes accuracy, precision, recall, and F1-score.
- Our approach achieved high accuracy compared to the most recent studies that used the same dataset.

The arrangement of the remaining sections of this paper is as follows: Section 2 discusses and summarizes related work. Section 3 provides a detailed explanation of the methods and materials employed in this study. Section 4 provides the results analysis, comparisons, and discussion. Section 5 presents the conclusion of this study.

## 2. RELATED WORK

The field of IoT security is attracting significant attention from researchers worldwide due to the rapid increase in IoT devices and a corresponding rise in security risks. This section is a concise summary of recent research focused on enhancing security in the IoT, specifically focusing on improving the security of the MQTT protocol in IoT environments. Table 1 briefly summarizes the relevant studies carried out in this domain.

Vaccari et al. [16] present the growing security concerns in IoT networks, explicitly introducing MQTTset, a dataset designed to study the MQTT protocol. The dataset contains malicious and legitimate traffic to detect attacks in the dataset, which can be utilized to train ML to identify IoT systems environments. The researchers highlight the significance of using a balanced dataset to get accurate detection attacks and properly evaluate algorithm performance. They applied different types of ML, such as RF, Gradient Boosting (GB), Neural Network (NN), Naive Bias (NB), DT, and multilayer perceptron (MLP), and the results demonstrated high accuracy and F1 scores.

Q. R. S. Fitni and K. Ramli et al. [17] propose an IDS by utilizing an ensemble learning approach to address data security problems in organizational information systems. The study chose three types of ML: decision trees, gradient boosting, and logistics regression as appropriate choices for the ensemble model and considered three types of attacks: brute force, SQL, and DoS. The authors removed 23 out of 80 features from the CSE-CIC-IDS2018 dataset, and the ensemble method technique enhanced IDS performance and achieved a final accuracy of 98.8% and an F1 score of 97.9%. However, the study proposed only uses the voting method for ensemble learning, so this study should be compared with other approaches, such as stacking, boosting, and bagging, to validate the study's findings and help guide future research.

Ahmad et al. [18] focus on using ML techniques for intrusion detection networks, which remarkably suggest using feature clusters based on Flow, MQTT, and TCP on the UNSW-NB15 dataset. The article discusses several classifications of methods, such as pre-processing and comparison with other approaches. The authors applied three types of ML: RF, Support Vector Machine (SVM), and Artificial Neural Networks (ANNs) for all the features. The study's findings demonstrated that the RF method achieved higher accuracy than the others for binary and multi-class classification. Nevertheless, this article needs to comprehensively analyze how the approach impacts the achievement of high performance.

F. Mosaiyebzadeh et al. [19] discuss the effectiveness of using a DL-based network IDS that is trained to identify MQTT attacks on the public MQTT-IoT-IDS2020 dataset. The authors used three DL methods, Convolutional Neural Networks (CNN), Recurrent Neural Networks (RNN), and long short-term memory (LSTM), to detect and handle different attack scenarios: aggressive scan, UDP-scan, and MQTT brute-force attacks. The DL approach is evaluated using multiple performance metrics such as precision, accuracy, F1-score, weighted average, and recall. The results of this evaluation achieved an F1 score of 98.33% and 97.09% accuracy. However, the authors need to go into greater detail to explain evaluation metrics.

Data sharing between devices is increasing with the fast growth of IoT networks. This is causing challenges due to the lack of sufficient security. Siddharthan et al. [20] introduced the IDS of elite machine learning algorithms (EML) for MQTT IoT networks. The authors created the EN-MQTTSET dataset, and they applied and evaluated several ML algorithms, including KNN, NB, GB, DT, Logistic Regression (LR), SVM, and RF. The result of the proposed method achieved over 99% accuracy in detecting intrusions in MQTT protocol attacks. However, this study needs more training to validate the system's effectiveness with different approaches.

To address the increasing security challenges that arise from the evolution of the number of IoT devices, Vijayan et al. [21] implemented an IDS approach to identify and categorize cyber-attacks. The researchers propose the CatBoost method to detect attacks for the MQTT protocol. This algorithm is trained on the MQTT network dataset that is publicly available, and it addresses the issue of imbalanced data. CatBoost has proven that this approach is superior to traditional ML methods such as MLP, NB, DT, RF, GB, NN, and NB. This method delivered a maximum accuracy of 94% within 78.45 seconds. However, this study needs to provide more details about the implementation and computational requirements of the CatBoost algorithm.

Akbar et al. [22] aim to detect brute force attacks on the MQTT protocol in IoT environments. The authors used the RF algorithm for data classification, which can manage large datasets and reduce the prevention of overfitting. They utilized two datasets: primary data from a hacking environment lab and the MQTT-IOT-IDS2020 dataset. The result of this study for both datasets indicated high results in accuracy, precision, recall, and f-measure. However, this study needs to compare with other methods and discuss its drawbacks and limitations.

Zeghida, H., Boulaiche, et al. [23] examine the security challenges of the MQTT protocol, particularly the detection of DoS attacks in IoT environments. The researchers employed DL techniques such as LSTM, CNN, GRU, CNN-RNN, CNN-LSTM, and CNN-GRU on the MQTT dataset to detect malicious behavior in IoT devices. This proposal's finding demonstrates high accuracy that exceeds 99% accuracy and a 98% F1-score, which is superior to traditional ML methods. However, this paper needs to provide more details regarding the preprocessing dataset.

Zeghida et al. [24] propose ensemble learning for the MQTT protocol to address the security challenges of IoT devices and identify susceptibility to cyber-attacks. They employed three known ensemble learning methods: bagging, boosting, and stacking. The model is trained on the MQTTset dataset, which is a binary balanced MQTT dataset. The result of this study shows that the network IDS has been enhanced in terms of accuracy, F1-score, and Matthews's correlation coefficient. However, this paper needs to provide a more detailed discussion of potential approaches to enhance the ensemble learning models' interpretability.

This study aims to detect and improve the performance of DoS and brute force attacks on MQTT network traffic by using feature engineering techniques with optimized ML models and ensemble learning methods. Table 2 compares our scheme and recently proposed methods, including feature selection, data balancing, cross-validation, ensemble classifiers, and evaluation metrics.

Table 1. Summary of related work

| Paper | Year | Dataset | Ensemble Technique | Approach | Evaluation metric | Attacks | MQTT protocol |
|---|---|---|---|---|---|---|---|
| [16] | 2020 | MQTTset | no | NN, RF, NB, DT, GB, MLP. | Accuracy, F1-score | Bruteforce, DoS, Flood, Malformed, Slowite | ✓ |
| [17] | 2020 | CIC-IDS 2018 | yes | DT, LR GB | Accuracy Precision Recall F1 score | Brute force, SQl, DoS | × |
| [18] | 2021 | UNSW-NB15 | no | RF, SVM, and ANN | accuracy | DoS | ✓ |
| [19] | 2021 | MQTT-IoT-IDS2020 | no | DNN, CNN, and LSTM | Accuracy F1 score | Scan_A, UDP scan, Brute force | ✓ |
| [20] | 2022 | SEN-MQTTset | yes | LR, DT, SVM, NB, KNN, GB, RF | Accuracy, F1-score, MCC, FAR, ROC, ADR, APV, NPV, | normal, sub-scriber attack, and broker attack. DoS, brute force, flooding | ✓ |

| | | | | | | | attacks | |
|---|---|---|---|---|---|---|---|---|
| [21] | 2023 | MQTTset | yes | (NN), RF, NB, DT, (GB), MLP, CatBoost | accuracy, Precision, Recall, F measure. | | Bruteforce, DoS,Flood, Malformed, Slowite | ✓ |
| [22] | 2023 | primary data MQTT-IOT-IDS2020 | no | RF | accuracy, Precision, Recall, F measure. | | Bruteforce | ✓ |
| [23] | 2023 | MQTT dataset | no | LSTM, CNN, GRU, CNN-RNN, CNN-LSTM, CNN-GRU | Accuracy, F1-score, Model loss | | DoS | ✓ |
| [24] | 2023 | MQTTset | yes | DT, RF, NB, MLP,Bagging, AdaBoost, HistGradientboost, XGB Classifier , Stacking | accuracy, F1-score, and MCC | | Bruteforce, DoS,Flood, Malformed, Slowite | ✓ |

Table 2. Comparison of our proposal with recently proposed models.

| Paper | Year | Dataset | ML Models | Data Balancing | Feature Selection | Cross-Validation | Ensemble Classifiers | Evaluation Metric |
|---|---|---|---|---|---|---|---|---|
| [16] | 2020 | MQTTset | + | + | - | - | - | Accuracy, F1-score |
| [21] | 2023 | MQTTset | + | + | - | - | +(CatBoost) | Accuracy, F1-score |
| ours | 2024 | MQTTset | + | + | + | + | +(Stacking, Voting, Bagging) | Accuracy, Precision, Recall, F measure |

Table 2 uses "+" to indicate the use of a method as a strength, while it uses "−" to indicate the non-use of a technique as a weakness.

## 3. METHODS AND MATERIALS

### 3.1. Proposed Methodology

This study aims to enhance the detection of intrusion DoS and brute force attacks in an MQTT traffic IDS within an IoT environment. The approach utilizes the MQTT dataset for model training by employing effective feature engineering and ensemble learning techniques. Through analysis and comparison, the features demonstrating the highest effectiveness are identified, leading to improved model accuracy. Multiple machines learning methods, including Decision Trees (DT), Random Forest (RF), K-Nearest Neighbors (KNN), and XGBoost, are evaluated to determine the most accurate and applicable models. Three ensemble methods are employed to optimize the proposed model's performance to enhance the efficacy of DoS and brute force attack detection systems. This approach combines the strengths of multiple base models to achieve superior results. The study utilizes various evaluation metrics, including accuracy, precision, recall, and F1-score.

This study used a Jupyter Notebook provided by the Anaconda platform. The Jupyter Notebook is a widely used tool that provides an interactive computing platform to facilitate executable code for users. Python language was used in this scheme because it is considered the best choice for ML libraries, can effectively handle complex computations, and is stable across various tasks.

### 3.2. Workflow of the Supervised ML Classifiers

The first workflow of the supervised ML models for the MQTT traffic IDS is illustrated in Figure 1. The first step is to import the dataset. The dataset was preprocessed by removing specific target values and unnecessary features. Then, we performed dataset preprocessing involving label encoding, data splitting, data normalization, and data balancing. Additionally, we utilized feature selection methods such as K-Best feature, PCC, and PCA, to select the most informative features. Subsequently, multiple ML methods were trained using the preprocessed data. Finally, the classification models were evaluated using performance metrics.

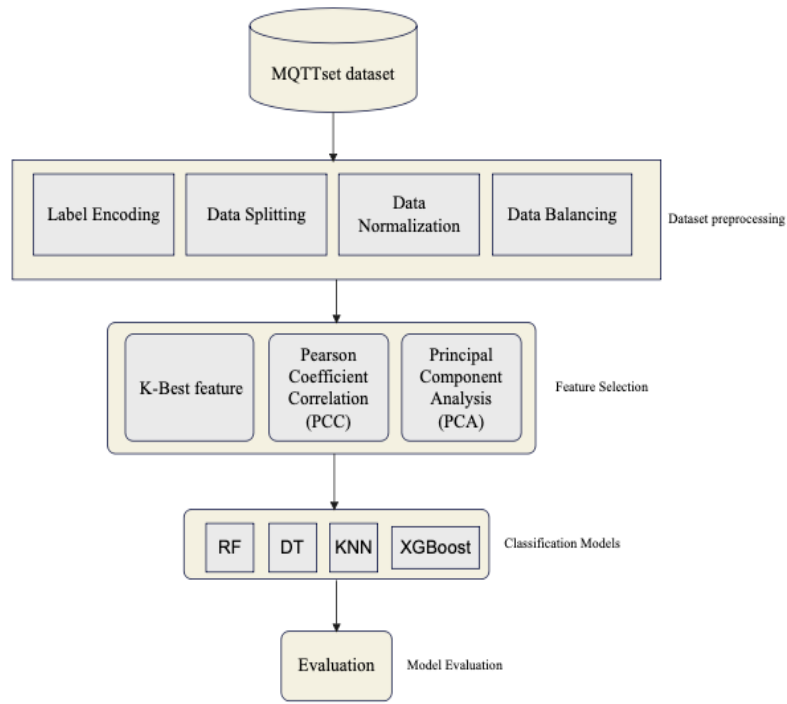

Figure 1. The workflow of MQTT network traffic in relation to the use of ML models.

## 3.3. Dataset Description

This research used the MQTTset dataset made available to the public domain in 2022 and obtained from Kaggle by Vaccari et al.[16]. This dataset was produced by using the MQTT protocol within an IoT smart home network, and the MQTT broker was constructed based on Eclipse Mosquitto. The network contains eight MQTT IoT sensors: CO-Gas, temperature, light intensity, smoke, fan, door lock, humidity, and motion sensor. These sensors are connected to an MQTT broker. All these sensors have varying time intervals because the specific functions of each sensor are different from the others. The dataset consists of 33 features, but we only used the ten most essential features based on the feature selection algorithms, as presented in Table 3. This dataset contains multiple attack scenarios, including legitimate and malicious; however, we focus on DoS and brute force attacks for this study.

## 3.4. Dataset Preprocessing

Data preprocessing is an essential phase of analyzing data and implementing ML methods. It enhances data quality, thereby improving the performance of ML algorithms and ensuring reliable data analysis. This step includes common techniques for handling missing data, encoding categorical variables, feature selection or extraction, normalization, and feature scaling. In this study, we utilized the reduced version of the MQTTset dataset available on the Kaggle platform by Vaccari et al. [16]. This dataset has been cleaned to address missing, extracted, and quality issues.

### 3.4.1. Label Encoding

Label encoding is an essential step in ML as it converts categorical features into numerical values. This study employed it due to its ability to transform features without increasing the number of features or computational complexity. We applied label encoding to several features of this dataset, including 'tcp.flags', 'mqtt.msg', 'mqtt.conack.flags', and 'mqtt.hdrflags'.

### 3.4.2. Data Splitting

Data splitting is an essential step in ML method development. It involves dividing the dataset into training and testing sets, where the training set is utilized to train the ML method, and the testing set is used to evaluate the model's performance on unseen data. This process ensures that it prevents overfitting and aids intrusion detection by enabling the identification of various types of attacks. This study divided the dataset into 80% training and 20% testing sets.

### 3.4.3. Data Normalization

Using data normalization in preprocessing data is essential. This technique is employed to normalize the range of features in the dataset. It also ensures that no single feature dominates the others during analysis and allows all features to contribute equally to the analysis [25]. This enhances the performance of the ML method and achieves a balanced influence on every feature during the analysis. This study applied Min-Max Scaling to scale the data to a consistent range, typically from 0 to 1.

### 3.4.4. Data Balancing

Class imbalance occurs when one class is much more common in classification than the others. To address this problem and create accurate and reliable classification models, techniques like the Synthetic Minority Over-sampling Technique (SMOTE) are commonly used [26]. In this study, we employed the SMOTE technique for oversampling. This approach involves generating synthetic instances for the minority class to balance the distribution of classes in the dataset. This enhances the accuracy of intrusion detection and mitigates the risk of overfitting.

### 3.5. Feature Engineering Methods

Feature engineering is an essential component to employ in the ML domain, as it enhances the model's performance by incorporating meaningful features and selecting the most relevant features from the dataset [27]. The selection of significant features in the dataset improves the IDS and achieves robust and more accurate ML methods. The feature section also offers several benefits, including reducing overfitting, faster training and inference, removing irrelevant or noisy features, concentrating on the most critical features, and reducing the model's complexity. This research used the feature selection algorithm of K-Best feature, PCC, and PCA, to select the best feature from the dataset.

After applying all three techniques of method selection, we chose the final set of the most repeated features that achieved a high score for each method. Table 3 presents the most important features for each method and the final set chosen.

International Journal of Artificial Intelligence and Applications (IJAIA), Vol.15, No.4, July 2024Table 3. Displays each technique's top features and the selected final set.

| Method for selecting features | Chosen features |
|---|---|
| K-Best feature | ['mqtt.msgid', 'tcp.len', 'mqtt.qos', 'mqtt.len', 'mqtt.hdrflags', 'mqtt.msg', 'mqtt.kalive', 'mqtt.msgtype', 'mqtt.conack.val', 'mqtt.conflag.uname'] |
| Pearson Coefficient Correlation (PCC) | ['mqtt.msgid', 'mqtt.qos', 'mqtt.len', 'tcp.time_delta', 'mqtt.msg', 'mqtt.hdrflags', 'mqtt.dupflag', 'tcp.len', 'tcp.flags', 'mqtt.conflag.cleansess', 'mqtt.proto_len', 'mqtt.protoname', 'mqtt.ver', 'mqtt.conack.flags', 'mqtt.conflags', 'mqtt.conack.val', 'mqtt.conflag.uname', 'mqtt.conflag.passwd', 'mqtt.kalive', 'mqtt.retain', 'mqtt.msgtype', 'mqtt.conack.flags.reserved', 'mqtt.conack.flags.sp', 'mqtt.conflag.qos', 'mqtt.conflag.reserved', 'mqtt.conflag.retain', 'mqtt.conflag.willflag', 'mqtt.sub.qos', 'mqtt.suback.qos', 'mqtt.willmsg', 'mqtt.willmsg_len', 'mqtt.willtopic', 'mqtt.willtopic_len'] |
| Principal Component Analysis (PCA) | ['mqtt.conflag.cleansess', 'mqtt.len', 'tcp.flags', 'mqtt.conack.val', 'mqtt.kalive', 'mqtt.retain', 'tcp.time_delta', 'tcp.len', 'mqtt.dupflag', 'mqtt.msgid'] |
| Final set | [Tcp.flags, tcp.time_delta, tcp.len, mqtt.dupflag, mqtt.hdrflags, mqtt.len, mqtt.msg, mqtt.msgid, mqtt.qos, mqtt.conack.flags] |

## 3.6. Classification Models

Selecting the appropriate ML model is essential because some of them are unsuitable for application scenarios. Table 4 presents the strengths and weaknesses of the chosen models utilized in this research.

Table 4. Displays the strengths and weaknesses of the selected models

| Machine learning techniques | Strength | Weakness |
|---|---|---|
| Decision Tree (DT) [28] | -It is easy to interpret, understand, and visualize. It can handle data normalization and categorical data. -It does not require feature scaling and less computational time. | -It is susceptible to overfitting when it becomes excessively complex. -It is unstable when small changes happen or something is added to the data. |
| Random forest (RF) [29] | -It generally achieves high accuracy and robustness for the overfitting in training data. - It helps to identify the most essential features in the dataset and reduce the variation. | -It can consume more memory because of its natural structures and the storage of multiple trees. -It is sensitive to the output of the user's choices of hyperparameters. |
| K-Nearest | -It is straightforward to understand and | -It is considered a less efficient |

| Neighbors (KNN) [30] | execute. <br> -It can be utilized for classification and regression tasks. | performance for making predictions in real time while working with enormous datasets. <br> -If the dataset size is large, it would be hard to find the K nearest neighbors in a short time. |
|---|---|---|
| XGBoost [28] | -It can handle missing values, which reduces the need for vast preprocessing in the dataset. <br> -It can be utilized for classification and regression tasks. | -It is sensitive when tuning the hyperparameter. |

### 3.7. Ensemble Classifiers' Learning

Ensemble techniques are a widely adopted approach in ML, which combines several independent models to improve predictive performance [31]. Ensemble methods help achieve superior performance by combining the strengths and mitigating the weaknesses of several models. Multiple ensemble approaches can be employed in IDSs. Some of the most widely used ones are as follows:

### 3.7.1. Stacking

It combines predictions from various base models by utilizing a meta-model, which then produces the final prediction. This classifier seeks to enhance predictive performance by combining the strengths of different base models [32]. This study consists of two phases: The initial step involves four base learners: RF, DT, KNN, and XGBoost. The second step involves a meta-learner utilizing a single model (RF). The optimal way to combine individual models in the base learners with the meta-learner is determined through these two processes.

### 3.7.2. Voting

A voting classifier is a technique that can combine several individual classifiers and provide better results. This technique provides much more accurate results compared to a single base model. The voting classifier can choose the best option from its ML models using a voting mechanism to make the final prediction [33]. Voting has two types: hard voting and soft voting. Hard voting depends on the majority vote of each classifier, while soft voting depends on the average probabilities across all classifiers. This study utilized hard voting to combine RF, DT, KNN, and XGBoost for classification.

### 3.7.3. Bagging

Bagging ensemble classifiers aim to improve the predictive performance of individual ML algorithms. This classifier helps effectively reduce the variance of the base model and prevents overfitting to enhance predictive performance. The bagging classifier achieves this by training several samples of the same base learning model on various subsets of the training data [34]. In this study, we utilized bagging with the RF model, which can lead to better predictive performance and reduce variance.

### 3.7.4. Workflow of Ensemble Classifiers

We utilized the preprocessed dataset from the supervised models demonstrated in (Section 3.6). This second workflow presents the proposed ensemble method model: stacking, voting, and bagging, as shown in Figure 2. This study aims to improve the accuracy of detecting DoS and brute-force attacks in the MQTT traffic in IoT environments through ensemble classifier prediction approaches. Enhancing the detection accuracy of these attacks can be achieved by combining multiple ML models to collaborate on a single ensemble classifier prediction.

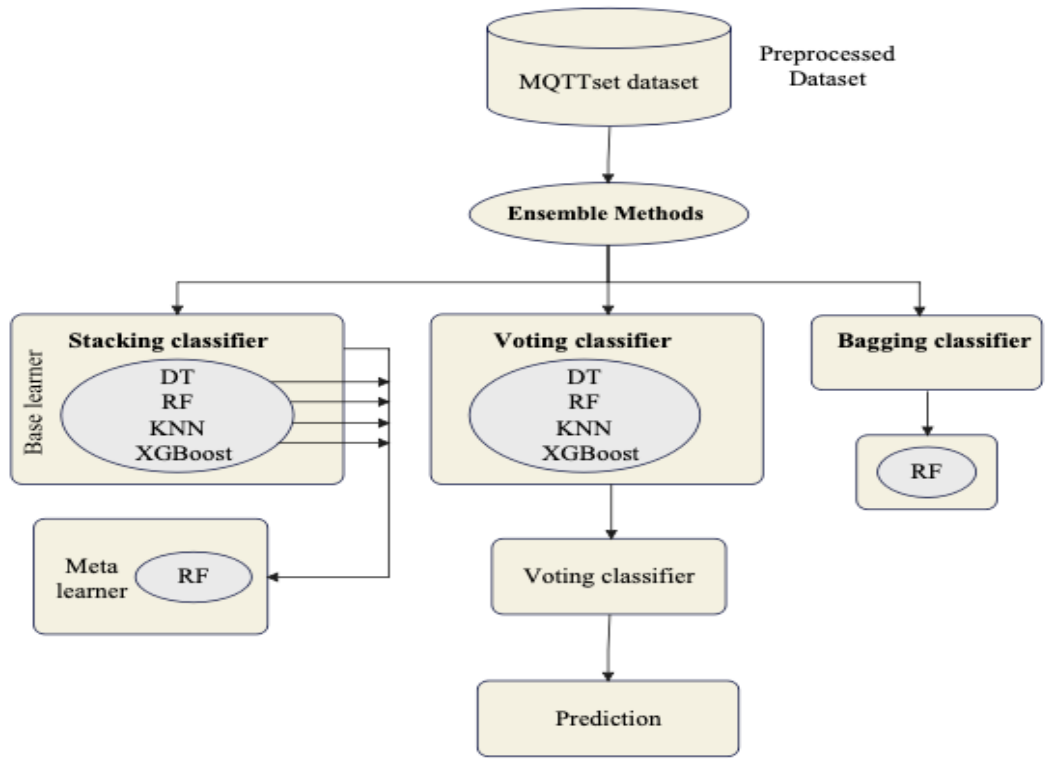

Figure 2. Depicting the workflow of the proposed ensemble classifiers.

## 4. RESULTS

### 4.1. Metrics for Evaluation

This study utilized multiple evaluation metrics, including accuracy, F1-Score, precision, and recall. These performance metrics assist in evaluating the effectiveness and reliability of DoS and brute force attack detection systems.

### 4.1.1. Accuracy

The model's accuracy is calculated as the proportion of correct prediction instances (TP and TN) to the total number of instances of predictions in the dataset (TP, TN, FP, and FN). The formula for accuracy is as follows [35]:

(1)    Accuracy = (TP + TN) / (TP + TN + FP + FN)

Where:

TP is a True Positive (correctly predicted positive samples)
TN is a True Negative (correctly predicted negative samples)
FP is a False Positive (incorrectly predicted positive samples)
FN is a False Negative (incorrectly predicted negative samples)

### 4.1.2. Precision

The model's precision is calculated as the ratio of correctly predicted positive samples (TP) to the total number of positive predictions made by the model in the dataset (TP and FP). The formula for precision is as follows [35]:

(2)     Precision= TP / (TP + FP)

### 4.1.3. Recall

The model's recall is calculated as the ratio of correctly predicted positive samples (TP) to the total number of actual positive samples (TP and FN). The formula for the recall is as follows [35]:

(3)     Recall= TP / (TP + FN).

### 4.1.4. F1-Score

The model's F1-Score is calculated as the weighted average of precision and recall utilizing the mean harmonic. The formula for F1-Score is as follows [35]:

(4)     F1-Score= 2 * (Precision * Recall) / (Precision + Recall).

## 4.2. Results and Discussion

This study aims to enhance the detection of DoS and brute force attacks in MQTT protocol traffic in IoT network systems. This section utilizes the dataset described in section 3-3 for experimental validation. As explained in section 3-4 on data preprocessing, several preprocessing techniques were performed, such as label encoding, data splitting, and data normalization. The (SMOTE) technique was utilized for oversampling to ensure an equal number of normal, DoS, and brute-force attacks in the dataset. Three different models, namely K-Best feature, PCC, and PCA, were employed to select the important features of the dataset that help improve the detection of DoS and brute-force attacks in MQTT traffic.

We employed efficient RF, DT, KNN, and XGBoost classifiers to detect normal, DoS, and brute-force attacks. The main objective is to combine them into ensemble methods such as stacking, voting, and bagging classifiers. The advantage of ensemble methods is that they improve the MQTT traffic IDS's performance in detecting DoS and brute force attacks. Each model evaluated the performance using accuracy, precision, recall, and F1-score across all the ML models and ensemble classifiers.

Furthermore, Table 5 displays the result of evaluating the supervised models to classify and detect normal, DoS, and brute force attacks. The finding in the provided table shows that each model can classify different kinds of attacks with high accuracy. The DT classifier achieved the highest accuracy at 0.9520, followed by the RF with 0.9519. They both excel in identifying DoS

International Journal of Artificial Intelligence and Applications (IJAIA), Vol.15, No.4, July 2024and brute force attacks. XGBoost classifier has achieved 0.9518 accuracy. The KNN classifier achieved a lower accuracy of 0.95. KNN classifier has a slightly lower precision in identifying DoS and brute force attacks than other algorithms while maintaining comparable recall and F1-score to different models. These results emphasize that choosing the most suitable model for particular application scenarios is essential to achieve high accuracy. Additionally, we chose RF, DT, KNN, and XGBoost classifiers for the proposed ensemble approaches due to their superior performance compared to other multiclass ML models.

Table 5. The evaluation results for DoS and brute force attacks using the supervised models.

| Methods | Attack | Precision | Recall | F1-Score | Accuracy |
|---|---|---|---|---|---|
| RF | Bruteforce: 0 | 0.93 | 0.88 | 0.90 | 0.9519 |
|  | Dos: 1 | 0.97 | 0.92 | 0.94 |  |
|  | Legitimate: 2 | 0.94 | 0.99 | 0.96 |  |
| DT | Bruteforce: 0 | 0.92 | 0.89 | 0.91 | 0.9520 |
|  | Dos: 1 | 0.97 | 0.92 | 0.94 |  |
|  | Legitimate: 2 | 0.94 | 0.99 | 0.96 |  |
| KNN | Bruteforce: 0 | 0.90 | 0.89 | 0.89 | 0.9500 |
|  | Dos: 1 | 0.96 | 0.92 | 0.94 |  |
|  | Legitimate: 2 | 0.95 | 0.98 | 0.96 |  |
| XGBoost | Bruteforce: 0 | 0.93 | 0.89 | 0.91 | 0.9518 |
|  | Dos: 1 | 0.97 | 0.91 | 0.94 |  |
|  | Legitimate: 2 | 0.93 | 0.99 | 0.95 |  |

In addition, our goal in this study is to enhance the detection of DoS and brute force attacks in MQTT traffic. We utilized four supervised ML models and combined them with ensemble classifiers. Table 6 presents the findings and evaluation of ensemble learning techniques, particularly stacking, voting, and bagging. These ensemble learning methods demonstrate the highest accuracy compared to single classifiers, as shown in Table 5. Ensemble learning achieved superior performance because it explored and enabled the collaboration of diverse learning mechanisms with different capabilities to improve each other's performance. The stacking classifier achieved high accuracy in identifying normal, DoS, and brute force attacks by 0.9538, with high results in precision, recall, and F1 scores. The process of stacking consists of two steps to optimize model performance. In the first step, we utilized four model-based classifiers: RF, DT, KNN, and XGBoost. In the second step, we used RF as the meta-learner for the stacking technique. Stacking enables the models to enhance each other's strengths and weaknesses using these two steps, eventually resulting in higher prediction performance.

The voting classifier achieved a high accuracy of 0.9538 to detect normal, DoS and brute force attacks. We utilized hard voting to combine RF, DT, KNN, and XGBoost during the prediction process.

It achieved high results for detecting DoS attacks for precision, recall, and F1-score at 0.97, 0.92, and 0.95, respectively. It also received high scores for detecting brute force attacks, with precision, recall, and F1 of 0.92, 0.90, and 0.91, respectively.

International Journal of Artificial Intelligence and Applications (IJAIA), Vol.15, No.4, July 2024Bagging classifiers achieved a higher accuracy of 0.9537 with high performance in terms of precision, recall, and F1 scores for normal, DoS, and brute-force attacks. We utilized a bagging classifier with the RF model during the prediction process.

This study was conducted multiple times, employing five-fold cross-validation to reveal components that may not be readily discernible during the initial training phase. This approach ensures the robustness and reliability of our findings.

In IoT environments, ensuring that IDS operates in real time is essential for mitigating and identifying attacks. As shown in Table 6, evaluations include training and testing times for all models, which help optimize model selection to maintain the security and responsiveness of IoT systems. Additionally, using multiple performance evaluation metrics such as accuracy, precision, recall, and F1-score helps ensure effective model operation and facilitates real-time improvements. Furthermore, while our proposed method enhances the detection of DoS and brute-force attacks within the MQTT environment, selecting optimal models is crucial to minimize potential countermeasures that could otherwise reduce its effectiveness.

Table 6. The evaluation results for DoS and brute force attacks using the ensemble methods.

| Methods | Attack | Precision | Recall | F1-Score | Accuracy | Training time | Test time |
|---|---|---|---|---|---|---|---|
| Stacking | Bruteforce: 0 | 0.93 | 0.90 | 0.91 | 0.9538 | 101.6051 | 8.5667 |
|  | Dos: 1 | 0.97 | 0.92 | 0.94 |  |  |  |
|  | Legitimate: 2 | 0.95 | 0.99 | 0.97 |  |  |  |
| Voting | Bruteforce: 0 | 0.92 | 0.90 | 0.91 | 0.9538 | 12.7990 | 7.4120 |
|  | Dos: 1 | 0.97 | 0.92 | 0.95 |  |  |  |
|  | Legitimate: 2 | 0.94 | 0.99 | 0.96 |  |  |  |
| Bagging | Bruteforce: 0 | 0.93 | 0.89 | 0.92 | 0.9537 | 76.2091 | 4.6069 |
|  | Dos: 1 | 0.97 | 0.92 | 0.94 |  |  |  |
|  | Legitimate: 2 | 0.94 | 0.99 | 0.96 |  |  |  |

The proposed IDS effectively detects normal, DoS, and brute force attacks and improves performance, as depicted in Figure 3, where the three proposed ensemble learning methods—stacking, voting, and bagging—outperform the four supervised models: RF, DT, KNN, and XGBoost.

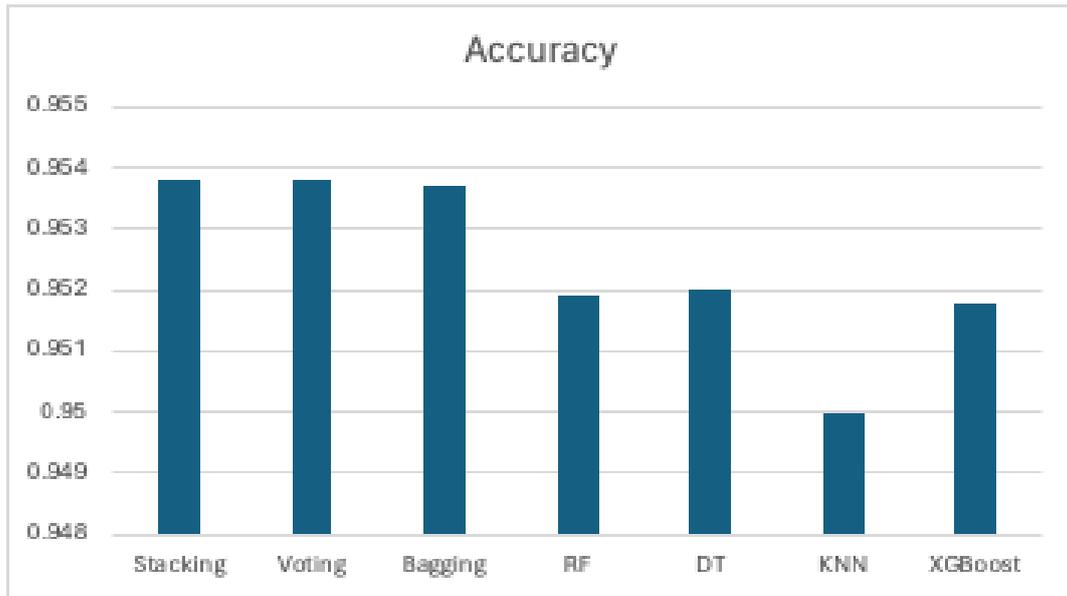

Figure 3. Comparing between ensemble classifiers and supervised methods.

The dataset used in our study is recent and extensively exploited in multiple studies aimed at identifying attacks in the MQTT environment. Table 7 displays the findings of our study compared to those of studies [16, 21], utilizing evaluation metrics for accuracy, precision, recall, and F1-score. The results of our study demonstrate the effectiveness of the proposed IDS MQTT traffic-based feature selection and ensemble classifier in enhancing the detection of DoS and brute-force attacks. Our evaluation findings show that the ensemble classifier achieved the highest accuracy in detecting all attacks. Paper [21] outperformed paper [16] in terms of accuracy for most ML techniques.

This study used a feature engineering technique, which enhances intrusion detection accuracy. We used three different models: K-Best feature, PCC, and PCA to select the most essential feature. We analyzed and compared all the features in the dataset through different steps. We then selected the top 10 features, as shown in Table 3, because they contribute to achieving high accuracy. These features have been used with an ensemble method to enhance the performance of MQTT traffic.

Papers [16, 21] employ all 33 features in the dataset, while our study uses the top ten feature selection. This indicates that papers [16, 21] did not utilize feature selection; therefore, our technique outperformed other studies in all evaluation metrics. Reducing feature selection helps remove irrelevant features and achieve high accuracy. Also, the studies of [16, 21] did not use cross-validation in their methodologies; on the other hand, our research utilized it, which helped ensure the robustness of the result. In addition, our proposed model depends on ensemble methods and feature engineering to enhance the detection of DoS and brute force attacks, thereby improving the dependability of the IDS for MQTT traffic. This approach considers thorough preprocessing, the optimal choice of ML models, data balancing techniques, and the application of cross-validation to ensure robust performance. These factors lead to multiple benefits, including improved accuracy, decreased dimensionality, and the faster testing and training of models.

Table 7. Comparing recently proposed models with our models using the same dataset.

| Paper | Year | dataset | Attacks | Models | Accuracy | Evaluation Metric | Feature selection | Cross-validation |
|---|---|---|---|---|---|---|---|---|
| [16] | 2020 | MQTTset | Brute force DoS Normal | DT | 0.9159 | Accuracy, F1-score | 33 | Not used |
| [21] | 2023 | MQTTset | Brute force DoS Normal | CatBoost | 0.9412 | Accuracy, F1-score | 33 | Not used |
| ours | 2024 | MQTTset | Brute force DoS Normal | Stacking Voting Bagging | 0.9538 0.9538 0.9537 | Accuracy, Precision, Recall, F1-score | 10 | used |

## 5. CONCLUSIONS

The MQTT protocol is widely utilized due to its lightweight and practical characteristics; however, it is susceptible to various security threats as it lacks built-in security mechanisms. In this study, we proposed an approach for enhancing the detection of DoS and brute force attacks on MQTT network traffic in an IoT environment. Our approach employs feature selection and integrates multiple traditional ML models into ensemble learning techniques to improve system performance. Feature selection enhances the efficacy and efficiency of cyber-attack categorization systems for MQTT traffic. Through our analysis of K-Best feature, Pearson Coefficient Correlation (PCC), and Principal Component Analysis (PCA), we identified the ten features that significantly impact the dataset. This selection has resulted in improvements in both the overall accuracy and speed of classification algorithms.

We examined multiple supervised ML models: RF, DT, KNN, and XGBoost. The findings demonstrated that DT achieved the highest accuracy at 0.9520, followed by the RF with 0.9519. Then, we combined RF, DT, KNN, and XGBoost by utilizing three ensemble techniques: stacking, voting, and bagging. The ensemble classifiers achieved better results than supervised ML models. Stacking and voting classifiers achieved high accuracy with 0.9538, while the bagging classifier achieved 0.9537. In each category of precision, recall, and F1 scores, all three ensemble classifiers demonstrated relatively similar levels of variability. The results indicate that utilizing feature selection with ensemble classifiers is crucial for improving the accuracy and reliability of categorization for DoS and brute force attacks on MQTT traffic. In addition, Table 7 shows that our ensemble classifiers, stacking, voting, and bagging results achieve superior accuracy compared to the models in [16,21]. In our future work, we aim to explore various types of attack detection methods and investigate more efficient and scalable approaches that have the potential to significantly enhance our results.